\documentclass[article,notoc]{JHEP3}
\usepackage{amsmath,epsfig}
\usepackage{amssymb,amsfonts}
\usepackage{latexsym}
\usepackage{graphicx}
\usepackage{epsfig}

\relax
\renewcommand{\theequation}{\arabic{section}.\arabic{equation}}
\renewcommand{\theequation}{\arabic{section}.\arabic{subsection}.\arabic{equation}}
\renewcommand{\theequation}{\arabic{section}.\arabic{equation}}

\def\be{\begin{equation}}
\def\ee{\end{equation}}
\def\bea{\begin{eqnarray}}
\def\eea{\end{eqnarray}}
\def\be{\begin{equation}}
\def\ee{\end{equation}}

\newcommand\fverb{\setbox\pippobox=\hbox\bgroup\verb}
\newcommand\fverbdo{\egroup\medskip\noindent%
                        \fbox{\unhbox\pippobox}\ }
\newcommand\fverbit{\egroup\item[\fbox{\unhbox\pippobox}]}
\newbox\pippobox
\def\F{\Phi}

\def\e{\epsilon}
\def\h{\eta}

\def\m{\mu}

\def\s{\sigma}
\def\th{\theta}
\def\t{\tau}

\def\p{\partial}

\def\f{\varphi}
\def\a{\alpha}

\def\g{\gamma}

\def\W{\Omega}
\def\la{\langle}
\def\ra{\rangle}

\def\ba{\begin{eqnarray}}
\def\ea{\end{eqnarray}}
\def\nb{\nonumber}
\def\jh{{\hat{\jmath}}}

\title{The symmetric branes of the $H_4$ WZW model}
\author{Giuseppe D'Appollonio\\
Department of Mathematics, King's College\\
The Strand, London WC2R 2LS, U.K.\\
{\tt E-mail: giuseppe@mth.kcl.ac.uk}}

\preprint{\hepth{0412258} \\ KCL-MTH-04-17}

\abstract{ We review the solution of the boundary CFTs that
describe the symmetric branes in the Nappi-Witten gravitational
wave, namely D2 and $S 1$ branes. The D2 branes are the twisted
branes of the model while the $S 1$ branes are the Cardy branes.
We present in both cases the bulk-boundary couplings and the
boundary three-point couplings and discuss the relation with
branes in $AdS_3$ and in $S^3$. We also discuss the analogy
between the open string couplings in the $H_4$ model and the
couplings for magnetized and intersecting branes. }

\begin{document}




\section{Introduction}

In the last few years important progress was made in the analysis
of some irrational CFTs. These theories are interesting on their
own and also relevant in order to describe the string dynamics in
curved space-times. Among other developments, the Liouville model
\cite{rliouville} and the $H_3^+$ model \cite{tads} were solved
and the structure of the string spectrum on $AdS_3$ clarified
\cite{moog1}. The study of string and brane dynamics in
non-compact curved backgrounds remains a rich and difficult
subject and all the models that can be solved exactly are a source
of useful information.

In \cite{dk} we started the analysis of a class of WZW models
based on the Heisenberg groups $H_{2n+2}$ \cite{nw,ks1,kehagias}
which from the space-time point of view describe the propagation
of a string in gravitational wave backgrounds. The gravitational
waves provide a rich family of conformally invariant $\s$-models
\cite{orig} and  a very convenient starting point for analyzing
the properties of string theory in curved, non-compact and
possibly singular backgrounds. They also play an important role in
the gauge/string duality. In fact the notion of Penrose limits
\cite{pen2} suggested that such backgrounds are dual to modified
large N limits of gauge theories \cite{mald}.

The models we are interested in describe a class of gravitational
waves that are supported by a NS-NS flux and have an exact CFT
description as WZW models. For more general gravitational wave
backgrounds one can investigate the spectrum of the model in the
light-cone gauge (for a recent discussion see \cite{papasolv}) but
a more detailed analysis of the string dynamics and in particular
the computation of the three and four-point amplitudes usually
remain beyond reach. On the other hand the presence of the affine
symmetry algebra allowed us to obtain a clear and detailed
understanding of the string dynamics in the gravitational waves
that correspond to the $H_{2n+2}$ WZW models. In \cite{dk}, we
solved the $H_4$ model without boundary and computed all the three
and four-point correlation functions of primary vertex operators.
In \cite{bdkz} we performed a similar analysis for the $H_{2n+2}$
models and in particular for the $H_{6}$ model, the Penrose limit
of $AdS_3 \times S^3$.

Here we will review the  D-branes of the Nappi-Witten
gravitational wave and we will discuss the dynamics of their open
string excitations \cite{dk2}. The dynamics of open strings in
curved space-time is even less understood than its closed
counterpart. Only the boundary states have been studied for many
of the interesting models that exist in the literature. The other
structure constants of the boundary CFT, namely the bulk-boundary
and the three-point boundary couplings, proved very difficult to
compute and only partial results are available at the moment
\cite{schads}. In \cite{dk2} we found the complete solution of the
BCFTs pertaining to the two classes of symmetric branes of the
$H_4$ model \cite{dbr,dbr2}, $D2$ and $S 1$ branes. We solved in
both cases the consistency BCFT conditions \cite{lew} and obtained
the complete BCFT data. To our knowledge, with the notable
exception of the Liouville model \cite{rliouville}, this is the
first {\em complete} tree-level solution of D-brane dynamics in a
curved non-compact background. Our solution of the $H_4$ model
with boundary should be useful not only to improve our
understanding of the closed and open string dynamics in curved
space-times but also to clarify some properties of both compact
and non-compact WZW models. In the following we will review, among
other results obtained  in \cite{dk2}, the first examples of
structure constants for twisted symmetric branes in a WZW model
(the D2 branes) and of open four-point functions in a curved
background. D-branes in pp-wave backgrounds have been the object
of several other studies and a review can be found in
\cite{dreviews}.

Before we start the analysis of the symmetric branes in the
Nappi-Witten gravitational wave, let us briefly describe two
additional reasons for paying attention to the $H_{2n+2}$ models,
besides the fact that they provide a rich and interesting class of
non-compact WZW models. The first reason is that, as explained in
\cite{ors}, the non-semi-simple WZW models can be considered as
contractions of semi-simple WZW models. If we assume this point of
view, we can implement at the worldsheet level the idea of Penrose
of realizing the gravitational waves as limits of other
space-times. Following \cite{dk,bdkz} we can study in detail how
the spectrum of the model and the dynamical quantities such as the
three-point functions change in the limit. As  far as the branes
are concerned, the fact that the $H_4$ model is the Penrose limit
of $\mathbb{R} \times S^3$ and $AdS_3 \times S^1$ leads to
interesting relations between the symmetric branes in $H_4$ and
the symmetric branes in $S^3$ and in $AdS_3$. The second reason is
that, according to the free-field representation introduced in
\cite{kk}, the $H_4$ primary vertex operators can be expressed
using orbifold twist fields. The orbifold in question is the
quotient of the plane by a rotation and a dictionary can be
established connecting the amplitudes computed in the $H_{4}$
model and the amplitudes computed in the orbifold CFT \cite{dk}.
Given this free-field representation, it is not surprising that in
the following we will find several analogies between the D2 branes
of the $H_4$ model and configurations of intersecting branes, as
well as between the $S 1$ branes of the $H_4$ model and branes
with a magnetic field on their world-volume.

\section{Branes in $H_4$ \label{a}}
\renewcommand{\theequation}{\arabic{section}.\arabic{equation}}

The $H_4$ WZW model was first analyzed as a string theory
background by Nappi and Witten \cite{nw}. They showed that the
target space of the corresponding $\s$-model is a gravitational
wave supported by a NS flux \be ds^2 = - 2 du dv - \frac{\m^2
r^2}{4} du^2 + dr^2 + r^2 d \f^2 \ , \hspace{1cm} H = \m r dr
\wedge d \f \wedge du \ . \label{a2} \ee The commutation relations
of the Heisenberg group are \be [P^+,P^-] = - 2i \m K \ ,
\hspace{1cm} [J,P^{\pm}] = \mp i \m P^{\pm} \ , \label{a3} \ee
where the generators $J$ and $K$ are anti-hermitian and
$(P^+)^{\dagger} = P^-$. The only subtlety that arises in the
construction of a WZW model based on a non-semi-simple group is
that in order to express the stress-energy tensor as a bilinear in
the currents one can not use the Killing form which is degenerate
but has to identify another non degenerate invariant symmetric
form. A detailed discussion of this model can be found in
\cite{dk} while a brief description of the same results can be
found in \cite{copenhagen}.

Since this gravitational wave is a WZW model, we can study in
considerable detail the symmetric branes, that is the branes that
preserve a linear combination of the left and right affine
algebras. The symmetric branes fall in families which are in
one-to-one correspondence with the automorphisms of the current
algebra and without loss of generality one can restricts his
attention to families of branes associated with distinct outer
automorphisms $\W$. The brane world-volumes coincide with the
twisted conjugacy classes of the group \cite{as,ffs}.

Since the $H_4$ algebra admits a non-trivial outer automorphism
$\Omega$ which acts on the currents as charge conjugation \be
\Omega(P^\pm) = P^\mp \ , \hspace{1cm} \Omega(J) = - J \ ,
\hspace{1cm} \Omega(K) = -K \ , \label{a08} \ee we have two
families of symmetric branes, wrapped on the conjugacy classes and
on the twisted conjugacy classes respectively \cite{dbr,dbr2}.

The $H_4$ conjugacy classes form a two parameters family ${\it
C}(u,\h)$, characterized by the constant value of the coordinate
$u$ and the constant value of the invariant \be \h = v - \frac{\m
r^2}{4} \cot{\frac{\m u}{2}} \ . \label{radius} \ee For
convenience we will often denote the two parameters that identify
an $S 1$ branes with a single letter $a \equiv (u_a,\h_a)$. Their
geometric description is particularly simple in Rosen coordinates
where the background takes the following form \be ds^2 = -2
dx^+dx^- +\sin^2 \frac{\m x^+}{2} (dy_1^2 + dy^2_2) \ ,
\hspace{1cm} H = \m \sin^2 \frac{\m x^+}{2} dy_1 \wedge dy_2
\wedge dx^+ \  . \label{a001} \ee Since in Rosen coordinates $x^-
= \h$, these branes are Euclidean two-planes with an
$x^+$-dependent scale factor and a two-form field-strength \be
{\cal F}_{12} \equiv B_{12} + 2 \pi \a^{'} F_{12} = - \frac{\sin
\m x^+}{2} \ , \label{sfluxr} \ee where as usual ${\cal F}$ is the
gauge invariant combination that appears in the Dirac-Born-Infeld
action. These branes have a non-trivial boundary condition on the
real time coordinate and can be called $S 1$-branes
\cite{S-branes}. In Brinkman coordinates, the metric induced on
the two-dimensional world-volume is trivial and the flux is \be
{\cal F}_{r\f} = B_{r \f} + 2 \pi \a^{'} F_{r \f} = - r \cot
\frac{\m u}{2} \ . \label{sfluxb} \ee When $\m u = 2 \pi n$, the
geometry of the conjugacy classes changes. We have either points
with  $r=0$ and a fixed value of $v$ or cylindrical branes
extended along the null direction $v$ and with a fixed radius  $r
\ne 0$ in the transverse plane. The relation between the $H_4$
vertex operators and the orbifold twist fields will lead to an
analogy between the $S 1$ branes in the Nappi-Witten wave and
branes with a magnetic field on their world-volume \cite{acny}.

The twisted conjugacy classes ${\cal C}^\W(\chi)$ are
parameterized by a single invariant \be \chi = r\cos{\f} \ , \ee
where we set $r e^{i \f} = \chi + i \xi$. In Brinkman coordinates
they have a simple description as $D2$ branes whose world-volume
covers the two light-cone directions and one direction in the
transverse plane. The induced metric is that of a pp-wave in one
dimension less and they also carry a null world-volume flux $F_{u
\xi} = \frac{\m \chi}{2}$. The $D2$ branes of the $H_4$ model thus
provide an interesting example of curved branes in a curved
space-time.
 The relation
between the $H_4$ vertex operators and the orbifold twist fields
will lead to an analogy between the D2 branes in the Nappi-Witten
wave and configurations of intersecting branes in flat space
\cite{angles}.

It is not difficult to identify how the symmetric branes in $H_4$
arise in the Penrose limit from the symmetric branes in
$\mathbb{R}\times S^3$ and $AdS_3 \times S^1$. In the first case,
the $S 1$ branes are the limit of $S^2$ branes in $S^3$ with a
Dirichlet boundary condition in the time direction and the $D2$
branes are the limit of rotated $S^2$ branes in $S^3$ with a
Neumann boundary condition in the time direction. In the second
case, the $S 1$ branes are the limit of the $H_2$ branes in
$AdS_3$ with a Dirichlet boundary condition in $S^1$ and the D2
branes are the limit of the $AdS_2$ branes in $AdS_3$ with a
Neumann boundary condition in $S^1$. A more detailed description
of the Penrose limit and of the contraction of the boundary
$\mathbb{R} \times SU(2)_k$ WZW model can be found in \cite{dk2}.

\section{The sewing constraints}

We now turn to the solution of the boundary CFTs that describe the
two classes of symmetric branes reviewed in the previous section.
For a CFT defined on the upper-half plane, there are two sets of
fields. The first set contains the bulk fields $\f_{i,\bar
\i}(z,\bar{z})$, inserted in the interior of the upper-half plane
and characterized by the quantum numbers $(i,\bar \i)$. These
quantum numbers
 specify the representations of the left and right chiral algebras. The
 second set contains the
 boundary  fields $\psi^{ab}_i(t)$, inserted on the boundary.
They are characterized by two boundary conditions $a$ and $b$.
They are also characterized by the quantum number $i$,  which
labels the representations of the linear combination of the left
and right affine algebras left unbroken by the boundary
conditions. The boundary conformal field theory is completely
specified by three sets of structure constants: the couplings
between three bulk or three boundary fields and the couplings
between one bulk and one boundary field. These structure constants
appear in three corresponding sets of OPEs \be \f_{i,\bar
\i}(z_1,\bar z_1)\f_{j,\bar \j}(z_2,\bar z_2) \sim \sum_{k}
(z_1-z_2)^{h_k-h_i-h_j} (\bar z_1-\bar z_2)^{h_{\bar k}-h_{\bar
\i}-h_{\bar \j}} C_{(i,\bar \i),(j,\bar \j)}{}{}{}^{(k,\bar k)}
\f_{k, \bar k}(z_2,\bar z_2) \ , \label{Obb} \ee \be \f_{i,\bar
\i}(t+iy) \sim \sum_{j} (2y)^{h_j-h_i-h_{\bar \i}}  \ {}^a
B_{(i,\bar \i)}^j \psi^{aa}_j(t) \ , \label{Occ} \ee \be
\psi_i^{ab}(t_1) \psi_j^{bc}(t_2) \sim \sum_{k}
(t_1-t_2)^{h_k-h_i-h_j} C^{abc,k}_{ij} \psi^{ac}_k(t_2) \ ,
\label{Obc} \ee where $t_1 < t_2$. The structure constants satisfy
a set of factorization constraints first derived by Cardy and
Lewellen \cite{lew}. These constraints involve, besides the
structure constants, the modular $S$ matrix and the fusing
matrices ${\bf F}_{p q}  \tiny
\begin{bmatrix} j & k
\\ i & l \end{bmatrix}$  that implement, by definition,
the duality transformations of the conformal blocks pertaining to
the four-point amplitudes.

In \cite{dk2} the bulk-boundary couplings ${}^a B^j_i$ and the
boundary three-point couplings $C^{abc}_{ijk}$ for the two classes
of symmetric branes of the $H_4$ WZW model were derived by solving
the sewing constraints. As an input we used the bulk three-point
couplings $ C_{(i,\bar \i),(j,\bar \j)}{}{}{}^{(k,\bar k)}$ and
the fusing matrices computed in \cite{dk}.

Before discussing the boundary CFT, we briefly recall the bulk
spectrum of the $H_4$ WZW model, whose structure is very similar
to the one established for $AdS_3$ \cite{moog1}. Together with the
standard highest-weight representations of the affine algebra,
restricted by a unitarity constraint, there are other
representations that satisfy a modified highest-weight condition.
These new representations are related to the standard ones by the
operation of spectral flow, which is an automorphism of the
current algebra. In our case the spectrum can be organized in
highest-weight and spectral-flowed representations of the affine
$\hat {\cal H}_4$ algebra and there are two distinct classes of
states. For generic values of the light-cone momentum $p$, the
states belong to the discrete representations of the $\hat {\cal
H}_4$ algebra. They correspond to short strings that are confined
by the background fields in closed orbits in the plane transverse
to the two light-cone coordinates. Whenever $\m \a\,' p \in
\mathbb{Z}$ the states belong to the continuous representations of
the $\hat {\cal H}_4$ algebra and correspond to long strings that
move freely in the transverse plane. A more detailed discussion of
these representations can be found in \cite{dk}. To each affine
representation we associate a primary chiral vertex operator \ba
&& \F^\pm_{\pm p,\jh}(z,x) \ , \hspace{1cm} 0 < \m p < 1 \ ,
\hspace{1cm} \jh \in \mathbb{R}
\ , \nb \\
&& \F^0_{s,\jh}(z,x) \ , \hspace{1.3cm} s > 0 \ ,  \hspace{1.9cm}
\jh \in [-\m/2,\m/2) \ . \ea For the $\F_{\pm p, \jh}^\pm$ vertex
operators, $p$ is the eigenvalue of $K$ and $\jh$ the highest
(lowest) eigenvalue of $J$. For the $\F_{s,\jh}^0$ vertex
operators, $s$ is related to the Casimir of the representation and
$\jh$ is the fractional part of the eigenvalues of $J$. Here $z$
is a coordinate on the world-sheet and $x$ a charge variable we
introduced to keep track of the infinite number of components of
the $H_4$ representations \cite{dk}. States with $\m p = \m \hat p
+ w$ with $0 < \m \hat p <1$ and $w \in \mathbb{N}$ fall into
spectral-flowed discrete representations $\Sigma_{\pm
w}(\F^\pm_{\pm \m \hat p, \jh})$ while states with $\m p = w$ with
$w \in \mathbb{Z}$ fall into spectral-flowed continuous
representations $\Sigma_{w}(\F^0_{s,\jh})$.

\section{The D2 branes}

The D2 branes of the $H_4$ model provide examples of curved branes
in a curved space-time and it is therefore very interesting to
study in  detail the dynamics of their open string excitations.
This is also important from a formal point of view. In fact the
couplings we discuss in this section are the first example of
structure constants for the twisted symmetric branes of a WZW
model and as such they could suggest how to extend to all possible
symmetric branes the general solution that is only available at
the moment for the untwisted branes.

The first thing we want to know is the spectrum of the open
strings ending on the D2 branes. The brane spectrum, exactly as
the bulk spectrum, can be organized in terms of highest weight and
spectral flowed $\hat {\cal H}_4$ representations. Not
surprisingly the open string spectrum of the D2 branes in $H_4$
has the same structure as the open string spectrum of the $AdS_2$
branes in $AdS_3$ \cite{oog1}. Also the spectral flow acts in the
same way, in particular spectral flow by an odd integer maps
strings whose ends are at positions $\chi_1$ and $\chi_2$ to
strings whose ends are at $\chi_1$ and $-\chi_2$. As a consequence
there is an asymmetry between the even and the odd spectral-flowed
continuous representations which is also manifest in the annulus
amplitude and in the analysis of the contraction of the boundary
$\mathbb{R} \times SU(2)_k$ WZW model \cite{dk2}. The spectrum of
a brane localized at $\chi \ne 0$ can then be summarized as
follows \ba && \Sigma_{2w} \left [ \psi^{\chi\chi}_{ \pm p,\jh}
\right ] \ , \hspace{0.8cm} 0 < \m p < 1 \ , \hspace{0.8cm} \jh
\in
\mathbb{R} \ , \hspace{0.8cm} w \in \mathbb{Z} \ , \nb \\
&& \Sigma_{2w} \left [ \psi^{\chi\chi}_{s,\jh} \right ] \ ,
\hspace{0.8cm}
 s \in \mathbb{R} \ , \hspace{0.8cm} \jh \in [-\m /2, \m/2  ) \ ,
 \hspace{0.8cm} w \in \mathbb{Z} \ , \nb \\
&& \Sigma_{2w+1} \left [ \psi^{-\chi\chi}_{s,\jh} \right ] \ ,
\hspace{0.8cm}
 s \in \mathbb{R} \ , \hspace{0.8cm} \jh \in [-\m /2, \m/2  ) \ ,
 \hspace{0.8cm} w \in \mathbb{Z} \ . \ea
The spectrum of open strings ending on two different branes
localized at $\chi_1$ and $\chi_2$ is essentially the same. The
only difference is that there is a minimal value for the quantum
number $s$ that reflects the tension of the string stretched
between the two branes.

We can now derive the structure constants by solving the sewing
constraints. The best way to proceed is to first derive the
bulk-boundary couplings and then the boundary three-point
couplings for strings ending on the D2 brane at $\chi=0$. Once
these couplings are obtained, the general solutions can be easily
identified.

The constraints for the bulk-boundary couplings follow from two
non-trivial bulk two-point functions, namely $\la \F^0_{s,\jh_1}
\F^0_{t,\jh_2} \ra$ and $\la \F^+_{p,\jh_1} \F^-_{-p,\jh_2} \ra$.
The first one, which is very similar to the corresponding
amplitude in flat space fixes ${}^\chi B_{s,\jh_1}^{(t,\jh_2)}$.
The second constraint then gives ${}^\chi B_{(\pm p,\jh)}^{(\pm
2p,2\jh_1 \pm n)}$. The bulk-boundary couplings with the identity
are particular cases of the previous couplings and they can be
used for the construction of the boundary states for the D2 branes
\cite{dk2,hikida}. Since the D2 branes are invariant under
translations along the two light-cone directions, only operators
with $p=0$ and $\jh=0, 1/2$ couple to their world-volume.

We now proceed to the computation of the boundary three-point
couplings. For the twisted branes of a WZW model there is no
general solution available and we have to solve directly the
constraints \cite{dk2}. The couplings between open strings living
on the brane at $ \chi=0$ are very similar to the square root of
the bulk couplings. For the complete solution we refer the reader
to \cite{dk2}. Here we display only a particular coupling that
nicely illustrate the relation between the D2 branes and
configurations of branes at angles \ba C^{\chi_1 \chi_2 \chi_3,
(p+q,\jh_1+\jh_2+n)}_{(p,\jh_1),(q,\jh_2)} &=&
\frac{2^{\frac{1}{2}} \pi ^{\frac{1}{4}} i^n}{2^{\frac{n}{2}}n!}
\left [ \frac{\g(p+q)}{\g(p)\g(q)} \right
]^{\frac{n}{2}+\frac{1}{4}}
e^{-\frac{Q^2}{2}} H_n(-Q) \ , \nb \\
 Q &=& \frac{(\chi_1 \sin \pi q - \chi_2
\sin \pi (p+q) + \chi_3 \sin \pi p)} {\sqrt{2 \pi \sin \pi p \sin
\pi (p+q) \sin \pi q}} \ . \label{sd2qb} \ea Let us compare this
coupling  with the couplings for intersecting branes discussed in
\cite{cim,cvetic,lust}. The quantum part of the boundary
three-point couplings, which can be computed using orbifold twist
fields, coincides with $(\ref{sd2qb})$ if we set $n=Q=0$. The $Q$
dependent exponential in $(\ref{sd2qb})$ can be interpreted as the
contribution of a disc world-sheet instanton \be {\cal C}_{ijk}
\sim e^{-\frac{A_{ijk}}{2\pi}} \ , \label{winston} \ee where
$A_{ijk}$ is the area of the triangle formed by the three
intersecting branes. In order to reproduce $(\ref{sd2qb})$ one has
to identify the angles between the branes with the light-cone
momenta of the three open strings in the gravitational wave and
the distance of the branes from the origin with the label $\chi$
of the D2 branes.

\section{The $S 1$ branes}

The analysis of the $S 1$ branes is greatly simplified by the fact
that they are the Cardy branes of the $H_4$ model. As a
consequence there is a relation between the parameters that label
the conjugacy classes and the quantum numbers of the $\hat {\cal
H}_4$ representations \be \m u = \pm 2 \pi (\m p + w) \ ,
\hspace{1cm} 2\h = \pi (2 \jh \pm 2 p \mp 1) \ , \label{rel} \ee
where as usual $0 < \m p <1 $ and $w \in \mathbb{N}$. This
relation can be established either  by studying the annulus
amplitudes of the $H_4$ model or by taking the Penrose limit of
the boundary $\mathbb{R} \times SU(2)_k$ WZW model \cite{dk2}. We
can now associate to a brane with labels $(u_a,\h_a)$ the
parameters $p_a$, $\jh_a$ and $w_a$. This is very useful because
as it is the case for the Cardy branes in a RCFT, the spectrum of
the open strings $\psi^{ab}$ ending on the brane $b$ and the brane
$a$ can be read from the fusion product $\Sigma_{\pm w_b}[\F_{\pm
p_b,\jh_b}] \otimes \Sigma_{\mp w_a}[\F_{\mp p_a,-\jh_a}]$ of the
two corresponding chiral vertex operators.

Using this information it is not difficult to identify the
spectrum of the open strings stretched between two $S 1$ branes.
The open strings that end on the same brane belong to the
continuous representations of the $\hat{\cal H}_4$ algebra. The
open strings that end on two different $S 1$ branes with labels
$a$ and $b$ can belong to any of the highest-weight
representations of the $\hat {\cal H}_4$ algebra as well as to
their images under spectral flow. The precise representation
depends on the distance between the two branes along the $u$
direction, according to $(\ref{rel})$. For the $S 1$ branes, the
spectral-flowed representations appear whenever the distance along
the $u$ direction between two branes exceeds $\frac{2\pi}{\m}$ and
it maps a string stretched between two branes localized at $u_a$
and $u_b$ to a string stretched from $u_a$ to $u_b+2 \pi w$.

The bulk-boundary structure constants can be fixed by studying the
factorization of three kinds of bulk two-point functions, namely
$\la \F_p^+\F_q^+ \ra$, $\la \F_p^+\F_q^- \ra$ and $\la \F_p^+
\F_s^0 \ra$. The only non-vanishing couplings are $ {}^a B_{\pm
p,\jh}^s$ and ${}^a B_{s_1,\jh}^{s_2}$ and can be found in
\cite{dk2}. The bulk-boundary couplings with the identity provide
all the information necessary for the construction of the boundary
states for the $S 1$ branes \cite{dk2,hikida}. Since the $S 1$
branes brake the translational invariance in the light-cone
directions, all the discrete representations and the continuous
representations with $s=0$ have a non-vanishing one-point
function.

The explicit form of the three-point couplings for the $S 1$
branes is slightly more involved than the form of the couplings
for the $D2$ branes discussed in the previous section. Their
derivation is however simpler. In fact the three-point boundary
couplings for Cardy boundary conditions in a RCFT, due to the
one-to-one correspondence between the boundary labels and the
representations of the chiral algebra, can be expressed using the
fusing matrices, as it was first realized for the Virasoro minimal
models \cite{ingo} \be C^{abc,k}_{ij} \sim  {\bf F}_{\check{b}k}
\small
\begin{bmatrix} i & j \\ a & \check{c}
\end{bmatrix} \ .
\label{cf} \ee In \cite{dk2} we verified that this relation
remains valid also for the non-compact $H_4$ WZW model, although
it is not a RCFT. For the general solution we refer the reader to
\cite{dk2}. Here we only discuss a very simple coupling in order
to illustrate the relation between the $S 1$ branes and branes
with magnetic fields. Consider the following coupling between
three open strings that live on the same brane \be C^{aaa,r}_{t s}
= \frac{1}{\pi st \sin \th} e^{\frac{i \pi st \sin \th}{2 \tan \pi
p_a}} \ , \label{b-3000} \ee with $r^2 = s^2+t^2-2ts \cos \th$.
The phase in $(\ref{b-3000})$ can be easily understood if we
compare this coupling with the coupling between open tachyon
vertex operators on a two-torus with a magnetic field $B$
\cite{acny}. We introduce two free bosonic fields $X_1$ and $X_2$
subject to the boundary conditions \be \p_\s X_1 + F \p_\t X_2 = 0
\ , \hspace{1cm} \p_\s X_2 - F \p_\t X_1 = 0 \ , \label{scb40} \ee
where $F = 2 \pi \a^{'} B$. From the OPE between open tachyon
vertex operators
 \be e^{i \vec p \vec X(t_1)} \,  e^{i \vec q \vec
X(t_2)} \sim (t_1-t_2)^{ 2 \a^{'} \frac{\vec p \vec q}{1+F^2}} \,
e^{ i \frac{\th^{ij}}{2} p_i q_j}  \, e^{i (\vec p+ \vec q) \vec
X(t_2)} \ ,  \hspace{0.4cm} \th^{ij} = -\frac{2 \pi \a^{'}
F}{1+F^2} \e^{ij} \ ,\label{scb43} \ee we can derive $F(u) = -
\cot(\m u/2)$ which is the expected result according to
$(\ref{sfluxb})$. Note that the magnetic field vanishes for $u =
\pi + 2 \pi n$. This corresponds to Neumann boundary conditions on
$X_1$ and $X_2$. Changing the value of $u$ we get the mixed
Neumann-Dirichlet boundary conditions in $(\ref{scb40})$ until we
reach $u = 2 \pi n$, where the field-strength diverges and
therefore the boundary conditions become pure Dirichlet. In fact,
precisely for these values of the coordinate $u$, the
two-dimensional conjugacy classes degenerate to points. According
to the analogy with open strings in a magnetic field, the strings
that live on the brane world-volume belong to the continuous
representations since their ends are both subject to the same
magnetic field and therefore they behave as free strings. On the
other hand, the strings stretched between two different branes
feel generically different magnetic fields and the corresponding
vertex operators are twist fields or, in $H_4$ terminology, they
belong to the discrete representations.

\section{Four-point amplitudes and generalizations}

With all the structure constants now at our disposal, we can in
principle compute arbitrary correlation functions by sewing
together the basic one, two and three-point amplitudes. Actually
in \cite{dk2} we discussed only those disc amplitudes that can be
expressed in terms of the four-point conformal blocks, namely
amplitudes containing either two bulk fields $\la \F \F \ra$ or
one bulk and two boundary fields $\la \F \psi \psi \ra$ or four
boundary fields $\la \psi \psi \psi \psi \ra$. Since explicit
expressions for the $H_4$ conformal blocks are available
\cite{dk}, these amplitudes can be studied in great detail. We
would like to mention that using the relation between the $H_4$
primary vertex operators and the orbifold twist fields \cite{kk},
the open string amplitudes in the Nappi-Witten gravitational wave
are generating functions for the correlators of arbitrarily
excited boundary twist fields in models with branes at angles
\cite{cim, cvetic, lust} or with branes with a magnetic field on
the world-volume \cite{narain}.

There are two aspects of our work we think deserve further study.
The first is to perform a more detailed analysis of the four-point
amplitudes and the second to clarify the relation between the open
and closed string channel of the annulus amplitudes. There are
also several simple generalizations of our work. In particular one
could study the symmetric branes of the $H_{2+2n}$ WZW models,
which have larger outer automorphism groups, or the branes of the
supersymmetric WZW models.

\section{Acknowledgments}

I am very grateful to Elias Kiritsis for a very enjoyable
collaboration and to Costas Bachas and Volker Schomerus for
several discussions. I would like to thank the organizers  of the
RTN workshop in Kolymbari for their kind invitation and for giving
me the opportunity to present our results. The work reviewed here
was done while I was supported by an European Commission Marie
Curie Individual Postdoctoral Fellowship at the LPTHE, Paris,
contract HMPF CT 2002-01908. My research is now supported by the
PPARC grant PPA/G/O/2002/00475. I also acknowledge partial support
by the EU network HPRN-CT-2000-00122 and by the Excellence Grant
MEXT-CT-2003-509661.



\begin{thebibliography}{10}

\bibitem{rliouville} For a review and a complete list of references see:
J.~Teschner, {\it Liouville theory revisited}, Class.\ Quant.\
Grav.\ {\bf 18} (2001) R153, hep-th/0104158;
Y.~Nakayama, {\it Liouville field theory: A decade after the
revolution}, Int.\ J.\ Mod.\ Phys.\ A {\bf 19}, 2771 (2004),
hep-th/0402009.



\bibitem{tads}
J.~Teschner, {\it On structure constants and fusion rules in the
SL(2,C)/SU(2) WZNW  model}, Nucl.\ Phys.\ B {\bf 546} (1999) 390
[hep-th/9712256];
{\it Operator product expansion and factorization in the $H_3^+$
WZNW model}, Nucl.\ Phys.\ B {\bf 571} (2000) 555
[hep-th/9906215].

\bibitem{moog1}
J.~M.~Maldacena and H.~Ooguri, {\it Strings in AdS$_3$ and SL(2,R)
WZW model. I}, J.\ Math.\ Phys.\  {\bf 42} (2001) 2929.


\bibitem{dk}
G.~D'Appollonio and E.~Kiritsis, {\it String interactions in
gravitational wave backgrounds}, Nucl.\ Phys.\ B {\bf 674} (2003)
80 [hep-th/0305081].



\bibitem{nw}
C.~R.~Nappi and E.~Witten, {\it A WZW model based on a
nonsemisimple group}, Phys.\ Rev.\ Lett.\  {\bf 71} (1993) 3751.



\bibitem{ks1}
K. Sfetsos, {\it Gauged WZW models and nonAbelian duality}, Phys.\
Rev.\ D {\bf 50} (1994) 2784 [hep-th/9402031];

\bibitem{kehagias}
A.~A.~Kehagias and P.~A.~A.~Meessen, {\it Exact string background
from a WZW model based on the Heisenberg group}, Phys.\ Lett.\ B
{\bf 331} (1994) 77.

\bibitem{orig}R.~G\"uven,
{\it Plane Waves In Effective Field Theories Of Superstrings},
Phys.\ Lett.\ B {\bf 191} (1987) 275;
D.~Amati and C.~Klimcik, {\it Strings In A Shock Wave Background
And Generation Of Curved Geometry {}From Flat Space String
Theory}, Phys.\ Lett.\ B {\bf 210} (1988) 92; G.~T.~Horowitz and
A.~R.~Steif, {\it Space-Time Singularities In String Theory},
Phys.\ Rev.\ Lett.\ {\bf 64} (1990) 260;
{\it Strings In Strong Gravitational Fields}, Phys.\ Rev.\ D {\bf
42} (1990) 1950.

\bibitem{pen2} M.~Blau, J.~Figueroa-O'Farrill, C.~Hull and G.~Papadopoulos,
{\it Penrose limits and maximal supersymmetry}, Class.\ Quant.\
Grav.\ {\bf 19} (2002) L87 [hep-th/0201081];
{\it A new maximally supersymmetric background of IIB superstring
theory}, JHEP {\bf 0201}, 047 (2002) [hep-th/0110242].

\bibitem{mald}
D.~Berenstein, J.~M.~Maldacena and H.~Nastase, {\it Strings in
flat space and pp waves from N = 4 super Yang Mills}, JHEP {\bf
0204} (2002) 013 [hep-th/0202021].


\bibitem{papasolv}
G.~Papadopoulos, J.~G.~Russo and A.~A.~Tseytlin, {\it Solvable
model of strings in a time-dependent plane-wave background},
Class.\ Quant.\ Grav.\ {\bf 20} (2003) 969 [hep-th/0211289].


\bibitem{bdkz}
M.~Bianchi, G.~D'Appollonio, E.~Kiritsis and O.~Zapata, {\it
String amplitudes in the Hpp-wave limit of AdS$_3 \times S^3$},
JHEP {\bf 0404}, 074 (2004) [hep-th/0402004].

\bibitem{dk2}
G.~D'Appollonio and E.~Kiritsis, {\it D-branes and BCFT in
Hpp-wave backgrounds},  hep-th/0410269.

\bibitem{schads}
B.~Ponsot, V.~Schomerus and J.~Teschner, {\it Branes in the
Euclidean AdS$_3$}, JHEP {\bf 0202} (2002) 016 [hep-th/0112198].


\bibitem{dbr}
S.~Stanciu and A.~A.~Tseytlin, {\it D-branes in curved spacetime:
Nappi-Witten background}, JHEP {\bf 9806} (1998) 010.

\bibitem{dbr2} J.~M.~Figueroa-O'Farrill and S.~Stanciu, {\it More D-branes in the
Nappi-Witten background}, JHEP {\bf 0001} (2000) 024;
{\it Penrose limits of Lie branes and a Nappi-Witten braneworld},
JHEP {\bf 0306} (2003) 025.

\bibitem{lew}
J.~L.~Cardy and D.~C.~Lewellen, {\it Bulk and boundary operators
in conformal field theory}, Phys.\ Lett.\ B {\bf 259}, 274 (1991);
D.~C.~Lewellen, {\it Sewing constraints for conformal field
theories on surfaces with boundaries}, Nucl.\ Phys.\ B {\bf 372}
(1992) 654.


\bibitem{dreviews}
M.~R.~Gaberdiel and M.~B.~Green, {\it D-branes in a plane-wave
background}, hep-th/0212052;
K.~Skenderis and M.~Taylor, {\it An overview of branes in the
plane wave background}, Class.\ Quant.\ Grav.\  {\bf 20} (2003)
S567, hep-th/0301221.

\bibitem{ors}
D.~I.~Olive, E.~Rabinovici and A.~Schwimmer, {\it A Class of
string backgrounds as a semiclassical limit of WZW models}, Phys.\
Lett.\ B {\bf 321} (1994) 361 [hep-th/9311081].



\bibitem{kk}
E.~Kiritsis and C.~Kounnas, {\it String Propagation In
Gravitational Wave Backgrounds}, Phys.\ Lett.\ B {\bf 320} (1994)
264 [Addendum-ibid.\ B {\bf 325} (1994) 536];
E.~Kiritsis, C.~Kounnas and D.~L\"ust, {\it Superstring
gravitational wave backgrounds with space-time supersymmetry},
Phys.\ Lett.\ B {\bf 331} (1994) 321.

\bibitem{copenhagen}
G.~D'Appollonio, {\it Gravitational waves from WZW models},
Class.\ Quant.\ Grav.\  {\bf 21} (2004) S1329, hep-th/0312206.

\bibitem{as}
A.~Y.~Alekseev and V.~Schomerus, {\it D-branes in the WZW model},
Phys.\ Rev.\ D {\bf 60} (1999) 061901.

\bibitem{ffs}
G.~Felder, J.~Frohlich, J.~Fuchs and C.~Schweigert, {\it The
geometry of WZW branes}, J.\ Geom.\ Phys.\  {\bf 34} (2000) 162.


\bibitem{S-branes}
M.~Gutperle and A.~Strominger, {\it Spacelike branes}, JHEP {\bf
0204} (2002) 018


\bibitem{acny}
A.~Abouelsaood, C.~G.~.~Callan, C.~R.~Nappi and S.~A.~Yost, {\it
Open Strings In Background Gauge Fields}, Nucl.\ Phys.\ B {\bf
280} (1987) 599.

\bibitem{angles}
M.~Berkooz, M.~R.~Douglas and R.~G.~Leigh, {\it Branes
intersecting at angles} Nucl.\ Phys.\ B {\bf 480} (1996) 265
[hep-th/9606139].


\bibitem{oog1}
P.~Lee, H.~Ooguri, J.~w.~Park and J.~Tannenhauser, {\it Open
strings on AdS$_2$ branes}, Nucl.\ Phys.\ B {\bf 610} (2001) 3.


\bibitem{hikida}
Y.~Hikida, {\it Boundary states in the Nappi-Witten model},
hep-th/0409185.





\bibitem{cim}
D.~Cremades, L.~E.~Ibanez and F.~Marchesano, {\it Yukawa couplings
in intersecting D-brane models}, JHEP {\bf 0307} (2003) 038;
{\it Computing Yukawa couplings from magnetized extra dimensions},
hep-th/0404229.

\bibitem{cvetic}
M.~Cvetic and I.~Papadimitriou, {\it Conformal field theory
couplings for intersecting D-branes on orientifolds}, Phys.\ Rev.\
D {\bf 68} (2003) 046001.

\bibitem{lust}
D.~Lust, P.~Mayr, R.~Richter and S.~Stieberger, {\it Scattering of
gauge, matter, and moduli fields from intersecting branes},
hep-th/0404134.


\bibitem{ingo}
I.~Runkel, {\it Boundary structure constants for the A-series
Virasoro minimal models}, Nucl.\ Phys.\ B {\bf 549} (1999) 563.

\bibitem{narain}
E.~Gava, K.~S.~Narain and M.~H.~Sarmadi, {\it On the bound states
of p- and (p+2)-branes}, Nucl.\ Phys.\ B {\bf 504} (1997) 214
[hep-th/9704006].


\end{thebibliography}
\end{document}